\documentstyle[sprocl]{article}

\input{psfig.sty}
\newcommand{\ar}{\arrowvert}
\newcommand{\ra}{\rangle}
\newcommand{\la}{\langle}
\newcommand{\da}{\dagger}
\newcommand{\ov}{\overline}
\newcommand{\cd}{\! \cdot \!}

\begin{document}
\title{Many-Body Approach to Mesons, Hybrids
and Glueballs}
\author{Stephen R. Cotanch and Felipe J.
Llanes-Estrada}
\address{Department of Physics, North Carolina
State
University, Raleigh NC 27695-8202}
\maketitle

\abstracts{We represent QCD at the hadronic scale by means of an
effective Hamiltonian,
$H$, formulated in the Coulomb gauge.  As in the
Nambu-Jona-Lasinio model,
chiral symmetry is dynamically broken, however our
approach is
renormalizable and  also includes confinement
through a linear potential
with slope specified by lattice gauge theory.
We perform a comparative
study of alternative many-body 
techniques for approximately
diagonalizing
$H$:  BCS for the vacuum ground state; TDA and
RPA for the
excited hadron states. 
We adequately describe the experimental meson
and
lattice glueball  spectra and perform the first relativistic,
three
quasiparticle calculation
for  hybrid mesons. In general agreement with alternative theoretical
approaches, we predict the lightest hybrid states near but above 2 $GeV$,
indicating the two recently observed $J^{PC} = 1^{-+}$ exotics
at 1.4 and 1.6
$GeV$   are
of a different, perhaps four quark,  structure. We also detail a new isospin
dependent interaction from $q\overline{q}$ color octet annihilation
(analogous to ortho positronium) which splits I = 0
and I = 1
states.}
\section{Introduction and Model Hamiltonian in the
Coulomb Gauge}
In the past several years we have
implemented
\cite{ssjc96,robertson,gjc,flesrc} an
ambitious QCD program to
comprehensively investigate
hadron structure.  Central to this project is a
realistic
effective Hamiltonian which we develop from QCD through
renormalization  and
subsequently diagonalize using established many-body
techniques.
We are particularly interested in the signature QCD prediction
of exotic
states (hadrons with quantum
numbers not possible in pure
$q\overline{q}$ or $qqq$ models).     
Early evidence \cite{hybrid1} for
the existence of an exotic hadron with
$J^{P C}=1^{-+}$, denoted
$\hat{\rho}$,
has finally been confirmed with the recent BNL observation
\cite{hybridexp2} of two states at 1.4 and 1.6 $GeV$.  These
measurements, along with anticipated future hybrid experiments (COMPASS
at CERN and Hall D at JLab),
have motivated the  study detailed in this
paper.

The theoretical situation is not satisfactory: lattice
results
\cite{hybridlat1} indicate that the lightest reliable hybrid meson
mass
with  quantum numbers $1^{-+}$ should appear at
$(2.0\pm 0.2)$  MeV.
On the other hand, vintage bag model results
\cite{Bagmodelhybrids} and a
recent QCD sum rule calculation
\cite{Sumruleshybrids}  agree in
predicting
exotic masses, between 1.4 and 1.7 GeV.
The sum rule prediction
\cite{Sumruleshybrids} claims a rigorous $1.9 \ GeV$ upper bound
which is clearly below  the lattice result.
The lattice calculations,
however, do require extrapolating for the $u, d$
sector and are therefore less accurate for the light quark hadrons.
Hence,
further insight is definitely needed to determine if these exotic
states are predominantly $q\overline{q}g$  objects
or belong to other Fock space sectors, such as
$q\overline{q}q\overline{q}$.

Our many-body Hamiltonian approach,
which
has been successfully tested for both glueballs \cite{ssjc96}
and conventional mesons \cite{flesrc}, addresses this issue.
Before treating the 3-body hybrid meson spectrum
we present our model Hamiltonian
in detail and briefly summarize our previous results.
Our QCD inspired, effective Hamiltonian is formulated in the Coulomb
gauge \cite{TDLEE}

\begin{eqnarray} \label{Hamiltonian}
H = \int d {\bf x}
\Psi ^{\dagger} \nonumber
({\bf x}) (-i{\bf \alpha}\!\cdot \! {\nabla} +
\beta m )
\Psi({\bf x}) + Tr \int d{\bf x} ( {\bf \Pi}^a  \!\cdot \!  {\bf
\Pi}^a +
{\bf B}_{A}^a  \!\cdot \! {\bf B}_{A}^a )
\\ - \frac{1}{2} \int d
{\bf x} d {\bf y}
\rho^a({\bf x})V(\arrowvert {\bf x} -
{\bf y} \arrowvert)
\rho^a({\bf y})    \ ,
\end{eqnarray}
containing both quark, $\Psi$, and
gluon, ${\bf A}^a$, 
${\bf B}_{A}^a =  \nabla \times 
{\bf A}^a $, fields
with color density
\mbox{$\rho^a = \Psi^{\dagger} \frac
{\lambda ^a} {2}
\Psi + f^{abc} {\bf 
A}^b \cdot {\bf \Pi}^c$}.
We adopt the standard current quark mass, $m$,
values for the $u, d, s$ and
$c$ flavors;
$m_u = m_d = 5 \ MeV$, $m_s = 150
\ MeV$ and $m_c = 1200 \ MeV$.
The linear interaction, $V_L = \sigma r$, is
obtained from lattice measurements and
Regge  phenomenology yielding
$\sigma = 0.18 \ GeV^2$.
For certain observables we supplement this with the canonical
Coulomb potential $V_C = -\frac {\alpha_s} {r}$
($\alpha_s = \frac {g^2_s} {4
\pi}$ $\approx 0.2 -0.4$).
The normal mode field expansions are

\begin{eqnarray}
\label{colorfields2}
A^a_i({\bf{x}}) =
\int
\frac{d{\bf{k}}}{(2\pi)^3}
\frac{1}{\sqrt{2\omega_k}}[a^a_i({\bf{k}})
+ a^{a\dag}_i(-{\bf{k}})]
e^{i{\bf{k}}\cdot {\bf {x}}} \\
\nonumber
\Pi_i^a({\bf{x}}) = -i \int
\frac{d{\bf{k}}}{(2\pi)^3}
\sqrt{\frac{\omega_k}{2}}
[a_i^a({\bf{k}})-a^{a\dag}_
i(-{\bf{k}})]e^{i{\bf{k}}\cdot
{\bf{x}}}
\\
\nonumber
\Psi({\bf{x}})=\sum_{c \lambda}
\int
\frac{d{\bf{k}}}{(2\pi)^3}
\left[u_{c\lambda}({\bf{k}})b_{c\lambda}({\bf{k}})
+
v_{c\lambda}(-{\bf{k}})d^{\dagger}_{c\lambda}(-{\bf{k}}) \right]
e^{i{\bf{k}}
\cdot {\bf{x}}} \ ,
\end{eqnarray}
and with the Coulomb gauge
transversality condition,
${\bf k} \cd {\bf a}^a ({\bf k}) = 0$,
yields the
commutation relation

\begin{equation} \label{conmutation}
[a_i^a({\bf
k}),a_j^{b\da}({\bf q})] = \delta^{ab}
(2\pi)^3\delta^{(3)}({\bf k}\! -
{\bf q}) (\delta_{ij}-\hat{k}_i \hat{k}_j)
\ .
\end{equation}
We note the
color density-density interaction, $\rho^a({\bf x}) V(\ar {\bf x}
\! -
\! {\bf y}
\ar ) \rho^a({\bf y})$, only produces color singlet states in
the spectrum
\cite{flesrc,Adler,Lisbon} when $V$ is a
linear, harmonic oscillator or any  Fourier
transformed confining interaction with a singular behavior
as $q
\longrightarrow k$.
This is a stringent check in our calculations, since
the matrix elements must always contain factors of $q-k$ to cancel this
singularity. In particular, for our chosen linear potential, 
we have in
momentum space  $\hat{V}(\ar 
{\bf q}-{\bf k} \ar)
=
-\frac{8\pi\sigma}{({q}-{k})^4}$.

\section {Ground State and BCS Mass Gap
Equations}\label{ground}

We now wish to solve $H \Psi = E \Psi$ as
accurately as possible.
In this section we focus on the ground state and
introduce the
Bardeen, Cooper and Schrieffer (BCS) transformation.
We
proceed by normal-ordering the Hamiltonian using the basis,
Eq. (\ref{colorfields2}), and
minimize the ground state, or vacuum,
expectation value.
The key concept in this approach is that of a quasiparticle. The bare,
current operators $a$, $b$, $d$ are rotated to improved
quasiparticle
operators $\alpha$, $B$, $D$ by
means of the BCS transformation. The vacuum
state, $\ar \Omega \ra$, is
determined by the relations $B \ar \Omega \ra = D \ar \Omega \ra = \alpha
\ar \Omega \ra =0$. 
The gluon and quark BCS rotations
are

\begin{eqnarray} \label{BVgluerotation}
 \alpha^a_i({\bf{k}}) = \cosh
\Theta_k \,
a^a_i({\bf{k}}) +
\sinh \Theta_k \, a_i^{a\da}(-{\bf{k}})
\\
\nonumber
B_{c \lambda }({\bf{k}}) = \cos \frac{\theta_k}{2} b_{c \lambda
}({\bf{k}})
 -\lambda \sin
\frac{\theta_k}{2} d_{c \lambda
}^{\dagger}({-{\bf{k}}})
\\ \nonumber
D_{c \lambda }(-{\bf{k}}) = \cos
\frac{\theta_k}{2} d_{c \lambda }(-{\bf{k}})
 +\lambda\sin
\frac{\theta_k}{2} b^{\dagger}_{c \lambda }({\bf{k}})  \
,
\end{eqnarray}
where $\Theta_k$, $ \theta_k/2 $ are the BCS angles,
further specified below.
This also introduces a redefinition of the
spectral function, $\omega_k$,
and counter-rotated quasiparticle spinors
$U$, $V$ 
such that the expansions of the fields remain formally invariant.

The specific variational parameters are the quark
gap angle, $\phi_k$,
related to the BCS angle by $tan (\phi_k - \theta_k) =
m /k$, and the gluon
self-energy, $\omega_k$, satisfying $\omega_k = k
e^{-2\Theta_k}$. The
vacuum contains correlated Cooper pairs
and  explicitly breaks chiral symmetry due to the  form of
$\omega$ and
$\phi$ \cite{flesrc,Lisbon,Orsay}.
The functional variation with respect to the two
parameters $\theta_k$,
$\Theta_k$
$$\delta \left(\frac{\la \Omega \ar H \ar
\Omega \ra}{\la \Omega
\ar
\Omega \ra} \right) = 0$$
generates two integral (mass gap) equations
for the
quark and gluon sectors,
respectively

\begin{equation}
\label{gap3d}
k\sin\phi_k -m\cos\phi_k =
\frac{2}{3} \int \frac{d{\bf
q}}{(2\pi)^3}
\hat{V}(\arrowvert {\bf k}
-{\bf
q} \arrowvert ) [\sin \phi_k \cos \phi_q \, \hat{{\bf k}} \cd
\hat{{\bf q}}
- \sin \phi_q \cos \phi_k] \end{equation}

\begin{equation}
\label{gluongap3d}
\omega_k^2 = k^2 - \frac{3}{4} \int
\frac{d{\bf
q}}{(2\pi)^3}
\hat{V}(\arrowvert {\bf k}
-{\bf q} \arrowvert )(1+ (\hat{{\bf k}}\cd \hat{{\bf q}})^2) \left(
\frac{w_q^2 - w_k^2}
{w_q} \right) \ .
\end{equation}
Significantly, the
BCS
vacuum is stable against quasiparticle pair creation since
the "anomalous" terms in the
Hamiltonian of the type $\alpha\alpha$,
$\alpha^\da \alpha^\da$, $BD$,
$B^\da D^\da$ are also cancelled by the same
rotation.

For the linear potential $1/q^4$, simple dimensional analysis
reveals
Eq. (\ref {gap3d}) is UV finite while Eq. (\ref {gluongap3d}) is
logarithmically
divergent. Hence we impose a momentum cutoff $\Lambda= 4
\ GeV $  and proceed
to solve both numerically after performing a standard three
dimensional
reduction. The details of this calculation are given
elsewhere
\cite{flesrc}.
We verify that our variational solution has a
minimum in
energy (traditional Mexican hat
shape).

\section{TDA and RPA for Mesons and
Glueballs}
\begin{figure}
\begin{center}
\psfig{figure=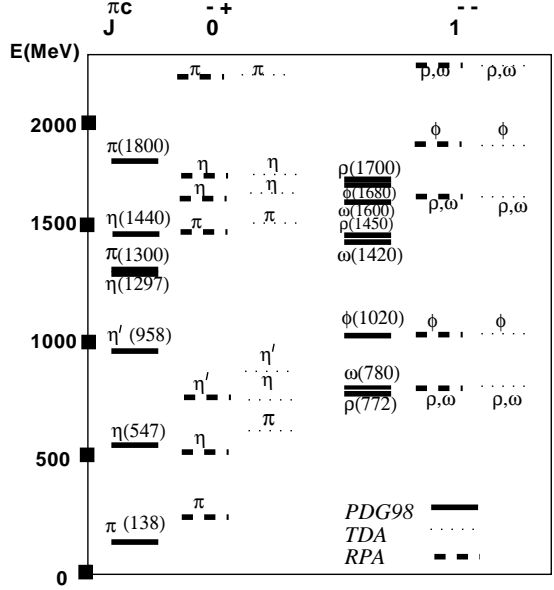,width=4in,height=2.7in}
\caption {TDA (dots), RPA (dashes) and
data (bars) for the light pseudoscalar and vector meson spectra.}
\label{meson}
\end{center}
\end{figure}

With these quasiparticle degrees of freedom we now construct
the
excited states from this vacuum.  For both the quark and gluon sector
we
represent hadrons as quasiparticle pairs (eg. $q\overline{q}$ for mesons
and $gg$ for glueballs) and angular momenta couple  to form  states of
good  $J^{PC}$. Next we invoke the TDA at
the 1p-1h level and diagonalize
the model Hamiltonian in this truncated
space.
A subset of our extensive TDA calculations is  displayed in
Fig. \ref{meson} for the pseudoscalar
mesons.  In general there is broad
agreement with the  data, the  most
notable exception being for the $\pi$ and
$\eta$.

The insufficient
$\approx 200\ MeV$, $\pi$/ $ \rho$ mass splitting
and related issue of
chiral symmetry motivated our improved RPA treatment.
Now the pion creation
operator generalizes to
$$
Q^\dagger = \sum_{ij}
\left( X_{ij}q^\dagger_i
\overline{q}^\dagger_j -
Y_{ij}q_i\overline{q}_j
\right),
$$
with pion state $\arrowvert \pi \rangle = Q^{\dagger} \arrowvert RPA \rangle$
and improved vacuum satisfying $Q \arrowvert RPA \rangle =0 $. Here $q_i $,
$\overline{q}_i$ are again the BCS rotated
quasiparticle operators and
$X_{ij}$, $Y_{ij}$ are the RPA
wavefunction components obtained from the
coupled equations of
motions generated by 
\begin{equation}
\label{RPA}
\langle \pi
\arrowvert [H,Q^\dagger] \arrowvert RPA \rangle =
M_{\pi} \langle \pi
\arrowvert  Q^\dagger \arrowvert RPA \rangle \ .
\end{equation}

In the chiral limit ($m=0$) the chiral condensate operator commutes
with $Q^\dagger$ and we rigorously compute $M_{\pi} = 0$
consistent with Goldstone's theorem.  For $m = 5 \ MeV$ explicit chiral
symmetry breaking yields $M_{\pi} = 294 \ MeV$, significantly better than 
the TDA value (note, the physical pion mass is reproduced with $m = 3
\  MeV$). Since both TDA and RPA produce comparable spin splittings, we
therefore conclude that chiral symmetry,
which only the RPA respects, is
responsible for most of the $\pi$/$\rho$ mass
difference.

Similarly, we generate the glueball spectrum for two quasiparticle 
gluons. Now chiral symmetry is not an issue and the RPA and TDA spectra 
agree to within a few percent. Our calculations, using the same string 
tension as above, reproduce  the lattice measurements as discussed in 
Confinement II.

A fundamental test for QCD is the existence of exotic
mesons (quantum numbers not possible in any
$q\overline{q}$  model).
In particular,  it has been speculated that two
recently observed
\cite{hybridexp2} states with isospin $1$ and  $J^{PC} =
1^{-+}$
at 1.4 and 1.6 $GeV$
contain explicit glue. Of several  gluonic
scenarios, glueballs
(oddballs) must be eliminated since they have isospin
$0$.  We therefore
investigate in the next section if these $1^{-+}$ states
are
hybrids.

\section{Three-Body TDA for Hybrid Mesons}
\label{3-body}

We formulate the hybrid meson as a 3-body
problem
($q\overline{q}g$) with TDA color wavefunction
$$\arrowvert
hybrid\rangle = B^{\dagger}
D^{\dagger}\alpha^{\dagger}\arrowvert
BCS
\rangle \equiv [[B^\dagger
\otimes D^\dagger]_8 \otimes
\alpha^\dagger]_0
\arrowvert BCS \rangle,$$
where the three BCS
quasiparticles have momenta ${\bf q}$, $\ov{{\bf q}}$,
and ${\bf g}$ referred to the hybrid $cm$ frame. The optimal choice of
relative 
variables is ${\bf q}_+= \frac{{\bf q}+\ov{{\bf q}}}{2}$,
${\bf q}_- = {\bf q}- \ov{{\bf q}}$ which facilitates implementing rotational,
parity ($P$) and  $C$-parity symmetries.
The complete TDA wavefunction
ansatz with  spin ($\lambda$) and color
($c, a$) indices
is
\begin{eqnarray} 
\nonumber
\ar hybrid \ra =\!
\sum_{\lambda_q \lambda_{\overline {q}}
\lambda c \overline{c} a}
\!\!\!\!
\int\!\!\!  \int \!\!  \frac{d{\bf q}_-}{(2\pi)^3}
\frac{d{\bf q}_+}{(2\pi)^3}
F^{J P C}_{\lambda_q \lambda_{\overline{q}} \lambda}({\bf q}_+,{\bf q}_-) 
T^a_{c \overline{c}} B^\da_{\lambda_q c}({\bf q}) D^\da_{
\lambda_{\overline{q}}
\overline{c}}({\bf \overline{q}})
\alpha_{\lambda}^{a\da} ({\bf g}) \ar \Omega \ra  \ .
\end{eqnarray}

This
wavefunction's  rotation properties are governed by the
three tensor indices $\lambda_q, \lambda_{\overline{q}}$,
$\lambda$, and the two arguments ${\bf q}_+, {\bf q}_-$ . To
construct an appropriate SU(2) representation we need to
couple five angular momenta (three quasiparticle intrinsic spins and two
orbital, $L\pm$, associated with ${\bf q} \pm$) to give a total
$J$, $m_J$.
To accomodate the Coulomb gauge transversality condition we utilize
a modified $LS$ coupling scheme involving two intermediate angular momenta
${\bf l} = {\bf L}_+ + {\bf 1}$ and ${\bf L} = {\bf l} + {\bf L}_-$. The
resulting angular momentum decomposition is 
\begin{eqnarray}
F^{J P
C}_{\lambda_g \lambda_q \lambda_{\overline{q}}}({\bf q}_+,{\bf q}_-)
=
\sum_{l L_- L_+ L S m_+ m_-}
F^{ J P C}_{l L_- L_+  LS}(\ar {\bf
q}_+\ar,\ar {\bf q}_-\ar) \, \,
Y^{m_+}_{L_+}(\hat{{\bf {q}}}_+) \,
Y^{m_-}_{L_-}(\hat{{\bf q}}_-)  
\nonumber \\
(-1)^{\lambda_g} \la L_+ m_+
1 -\lambda_g \ar l m_l \ra \la L_- m_- l m_l
\ar L m_L \ra 
\nonumber
\\
\la \frac{1}{2} \lambda_q \frac{1}{2} -\lambda_{\overline{q}} \ar S m_S
\ra (-1)^{\frac{1}{2} -
\lambda_{\overline{q}}} \la L m_L S m_S \ar J m_J
\ra \nonumber \ .
\end{eqnarray}
Due to the transversality condition
mentioned above, $\hat{\bf q}_+\cd
{\bf \alpha} =0 $,
so
\begin{equation}
\la 1 m_+ 1 \lambda \ar 0 0 \ra Y_1^{m_+}(\hat{\bf
q}_+) 
\alpha_{\lambda}^\da  =0
\end{equation}
and therefore we cannot
couple the angular momenta of the gluon
to give the intermediate state $\ar
l m_l \ra = \ar 0 0 \ra$.

With this coupling scheme the total parity is a
product of the
intrinsic parities of the quark-antiquark pair (-1), the
intrinsic parity of the gluon  (-1), and the two spherical
harmonics yielding

$$ P = (-1)\cd(-1)\cd(-1)^{L_+ + L_-} = (-1)^{L_+ + L_-} \ .$$

Similarly the total charge conjugation is a product of the
quark-antiquark
pair ($(-1)^{L_-+S}$ from
equivalence to the exchange of all
$q\overline{q}$
quantum numbers) and the gluon (-1), which is
its own
antiparticle and has odd C-parity. Hence

\begin{table}
\caption{Possible
Quantum Numbers for a $q\overline{q}g$ Hybrid Meson}
\label{qnumbers}
\vspace{0.3in}
\begin{center}
\begin{tabular}{|c|c|c|c|c
|c|c|c|}
\hline
  $L_+$  &  $L_-$  &  $S$  &  $L$  &  $J$  & $ P$  &  $C$ &
\\
\hline
  0 & 0 & 0 & 1 & 1 & + & - & \\
  0 & 0 & 1 & 1 & 0 & + & + &
\\
  0 & 0 & 1 & 1 & 1 & + & + & \\
  0 & 0 & 1 & 1 & 2 & + & + &
\\
\hline
  0 & 1 & 0 & 0 & 0 & - & + & \\
  0 & 1 & 0 & 1 & 1 & - & +
&
Exotic \\
  0 & 1 & 0 & 2 & 2 & - & + & \\  
  0 & 1 & 1 & 0 & 1 & - & -
&
\\
  0 & 1 & 1 & 1 & 0 & - & - & Exotic \\
  0 & 1 & 1 & 1 & 1 & - & -
&
\\
  0 & 1 & 1 & 1 & 2 & - & - & \\
  0 & 1 & 1 & 2 & 1 & - & - & \\
  0
& 1 & 1 & 2 & 2 & - & - & \\
  0 & 1 & 1 & 2 & 3 & - & - & \\
\hline
  1 &
0 & 0 & 0 & 0 & - & - & Forbidden by transversality\\
  1 & 0 & 0 & 1 & 1 &
- & - & \\
  1 & 0 & 0 & 2 & 2 & - & - & \\
  1 & 0 & 1 & 0 & 1 & - & + &
Forbidden by transversality\\
  1 & 0 & 1 & 1 & 0 & - & + & \\
  1 & 0 & 1
& 1 & 1 & - & + & Exotic \\
  1 & 0 & 1 & 1 & 2 & - & + & \\
  1 & 0 & 1 &
2 & 1 & - & + & Exotic \\
  1 & 0 & 1 & 2 & 2 & - & + & \\
  1 & 0 & 1 & 2
& 3 & - & + & Exotic \\
\hline
\end{tabular} \end{center}
\end{table}

$$ C = (-1)\cd(-1)^{L_- +S} = (-1)^{1+L_-
+S} \ .$$
Notice the extra C-parity sign which now permits
exotic quantum
numbers.
We are only interested in the lightest hybrids which will have S
and P waves
for either of the angular momenta $L_+$ or $L_-$. The
resulting
possible quantum numbers are displayed in
Table
\ref{qnumbers}.

In principle all the states with the same $J^{P C}$
quantum
numbers can mix but we do not address this issue here.
We also note
that our wavefunction normalization is not standard due to
the
transversality relation, Eq. (\ref{conmutation}),  which introduces
an
extra angular momentum
coefficient.

\begin{figure}
\psfig{figure=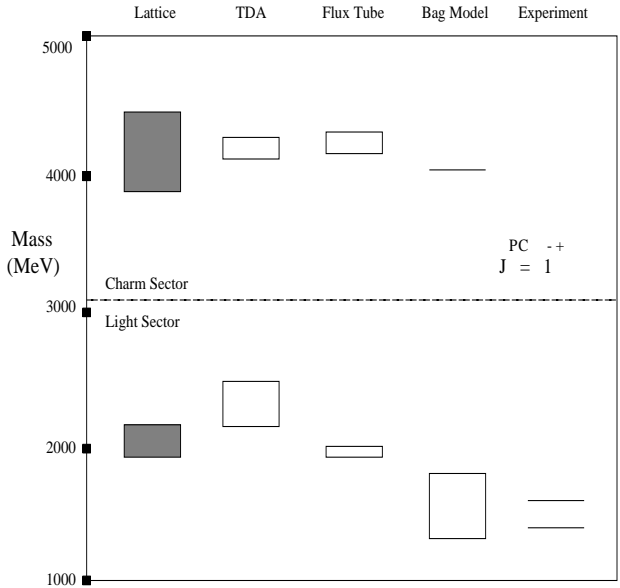,width=3.5in,height=6in}
\vspace{-7.5cm} \caption
{Light and charmed exotic $1^{-+}$ hybrid
mesons.
The TDA calculation is comparable to the lattice results but
both
disagree with observed resonances.}
\label{hybrids}
\end{figure}

Finally, in analogy with our two-body TDA treatment the hybrid equation
of motion is \begin{equation} \label{TDAhybrid}
\la q \ov{q} g \ar H \ar
hybrid \ra = E \la q \ov{q} g \ar hybrid \ra \ .
\end{equation}
Using Eqs.
(\ref{Hamiltonian},
\ref{colorfields2}) and the hybrid wavefunction
expansion, we obtain the
TDA equation of motion which now is a  non-local
equation in the two 
variables ${\bf q}_+$, ${\bf q}_-$
(12-dimensional
problem) precluding practical matrix diagonalization.
Therefore we utilize
alternative ans\"{a}tze for the radial wavefunction and
variationally
evaluate the Hamiltonian matrix element.

Separating the $cm$
variables,
the integrals can be reduced to 9-dimensions
which we evaluated by  Monte
Carlo methods utilizing Lepage's code 
\mbox{VEGAS}.
We employed an
increasing number of points, up to several million,  until
satisfactory
convergence was achieved.  The angular wavefunctions are
explicitly coded
in terms of spherical harmonics with all magnetic spin
sums and angular
integrals  evaluated numerically.
The integrable IR singularity from our
potential received special
attention.  We divided the complete integral
into different parts
depending upon the Hamiltonian components and then
used a change of variables,
always placing the singularity at the origin, with
a Jacobian transformation to concentrate points there. This is
detailed in
our previous work.

In Fig. \ref{hybrids}  we present the main result of
this work, our 
prediction of the
light and charmed exotic hybrid mass. We
also compare to alternative
lattice, flux
tube and even vintage bag model
approaches as well as the recent E852
Brookhaven data.
Notice that, with
the exception of the dated bag model results, the
different theoretical
models generally agree but disagree with observation.

Because of the
potential ramifications of this disagreement between
measurement
and
predictions we further examined and tested our model and calculation in
detail.
First, to ensure our variational method is accurate, we compared
several
different  radial trial wavefunctions. In particular, we
used
gaussian and also more complicated functions generated by the exact
numerical solutions to simpler, 2-body problems ($\rho$ meson,
$J/\Psi$,
glueball) with various potentials. We also added nodes as
needed. We found
the maximum variance (sensitivity)  to
wavefunction choice to be about 50
MeV in the hybrid mass.
Second, we also considered whether chiral symmetry
affects
the hybrid meson spectrum and variationally formulated the RPA as
above,
but now with an additional constituent TDA gluon.
Interestingly, we find
RPA and TDA to be numerically equivalent. The
reason  is that the quarks
are now in a color octet state but
the chiral octet charge
$$ Q_5^a = \int
d{\bf x} \Psi^\da({\bf x}) \gamma_5 T^a \Psi({\bf x}) $$
is not conserved
since the commutator with the Hamiltonian
does not vanish.
Further, there
is also no Gell-Mann-Oakes-Renner relation
and no Goldstone hybrid boson in
the chiral limit $m\longrightarrow 0$.
Therefore we do not expect any major
numerical effect in the 
hybrid spectrum, analogous to the 400 MeV downward
shift of the chiral pion.
This was confirmed by an explicit numerical
calculation in a selected 
channel.

Consequently, we are reasonably confident in our prediction that the 
lightest exotic hybrid mass is above 2 $GeV$ and concur with other
contemporary theoretical model results.  This also strongly
indicates that the observed exotic $1^{-+}$ at 1.4 and 1.6
$GeV$ are not hybrid mesons but have an alternative, four quark
or meson molecular structure.

Our last key result concerns the charmonium $J/\Psi$ spectrum and is shown
in Fig. 3. By simply calculating D wave TDA $c\overline{c}$
states we have resolved the historic "overpopulation"
problem.
Ironically, if we then include our predicted non-exotic charmonium states,
also shown in the figure, we now have an "underpopulation" problem
which is more typical of confrontation between the conventional quark model
and experiment.  It would now appear that simply counting
states may not be that effective in identifying gluon rich
states.

Finally, since any three quasiparticle problem can be treated
variationally, we have explicitly constructed valence baryons
($qqq$) states and evaluated both the nucleon ($N$) and
$\Delta_{3/2}$ with our same Hamiltonian.  Our $\Delta$ mass is in 
agreement with data, but only about $1/3$ of the $N$-$\Delta$ mass splitting
 is obtained.  The other interesting case is the triple gluon
glueball, $ggg$, whose study is still preliminary and will be reported in the
near future. This  exhausts the possible three body states which can form color
singlets.

\begin{figure}
\psfig{figure=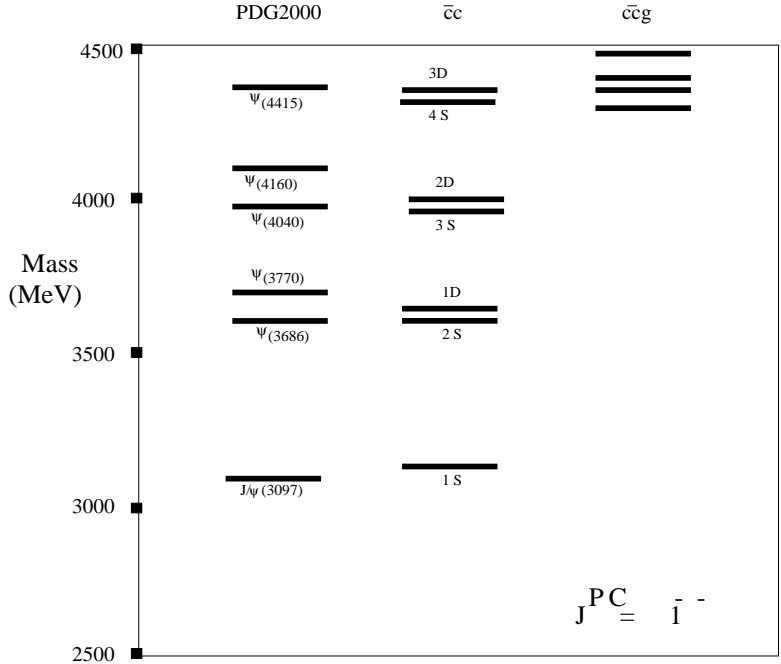,width=4.5in,height=6in}
\vspace{-2.5in}
\caption{TDA theory for conventional ($c\overline{c}$) and non-exotic 
($c\overline{c}g$) states compared to the observed
$1^{--}$ $J/\psi$ spectrum$^14$.}
\label{spectrum2}
\end{figure}

\section{Conclusions}
\label{3body}

In summary, the BCS, TDA and RPA many-body treatments are
powerful,
effective methods for investigating hadron structure.  Using a
relativistic, field theoretical Hamiltonian with standard quark masses
and only one (pre-determined) interaction parameter, our approximate
many-body solutions yield reasonable descriptions of the vacuum, mesons and
glueballs. In particular, the RPA properly incorporates
chiral symmetry and provides new insight into the condensate structure of
the vacuum and chiral governance of the pion.  Most significantly, our
hybrid meson mass prediction is above 2 $GeV$ and in reasonable agreement 
with both lattice and flux tube results. This strongly suggests that the 
recent observed exotic states below 2 $GeV$ are not hybrids but more 
likely 4 quark states.   Because our method is directly amenable to
including higher Fock state components, we are currently calculating the
molecular ($q\overline{q} q\overline{q}$) meson spectrum.  Future work
will also entail extensions to the nucleon strangeness content (pentaquark
systems) and even dibaryons (six quark systems) which are both tractable
in our approach.

\section*{Acknowledgments}
We would like to thank our collaborators P. Bicudo, E. Ribeiro, A.
Szczepaniak and the NCSU theory group. This work was partially
supported by grants  DOE DE-FG02-97ER41048 and NSF INT-9807009.
F. J. Llanes-Estrada is grateful for a SURA-Jefferson Laboratory graduate
fellowship. Supercomputer time from NERSC is also acknowledged.

\end{document}